\documentclass[letterpaper,USenglish]{lipics-v2018}
\hideLIPIcs
\pdfoutput=1

\title{Table Based Detection of Degenerate Predicates in Free Space Construction}

\author{Victor Milenkovic}{Department of Computer Science, University of Miami\\
  Coral Gables, FL 33124-4245, USA}{vjm@cs.miami.edu}{}{Supported by NSF CCF-1526335.}

\author{Elisha Sacks}{Computer Science Department, Purdue University\\
  West Lafayette, IN 47907-2066, USA}{eps@purdue.edu}{}{Sacks and Butt supported
  by NSF CCF-1524455.}

\author{Nabeel Butt}{Facebook\\
  1 Hacker Way, Menlo Park, CA 94025, USA}{nfb@fb.com}{}{}

\authorrunning{V. Milenkovic, E. Sacks, and N. Butt}

\Copyright{Victor Milenkovic, Elisha Sacks, and Nabeel Butt}

\subjclass{}


\EventEditors{Bettina Speckmann and Csaba D. T{\'o}th}
\EventNoEds{2}
\EventLongTitle{34th International Symposium on Computational Geometry (SoCG 2018)}
\EventShortTitle{SoCG 2018}
\EventAcronym{SoCG}
\EventYear{2018}
\EventDate{June 11--14, 2018}
\EventLocation{Budapest, Hungary}
\EventLogo{socg-logo} 
\SeriesVolume{99}
\ArticleNo{62} 

\nolinenumbers

\begin{document}

\maketitle

\begin{abstract}

The key to a robust and efficient implementation of a computational geometry
algorithm is an efficient algorithm for detecting degenerate predicates.  We
study degeneracy detection in constructing the free space of a polyhedron that
rotates around a fixed axis and translates freely relative to another
polyhedron.  The structure of the free space is determined by the signs of
univariate polynomials, called angle polynomials, whose
coefficients are polynomials in the coordinates of the vertices of the
polyhedra.  Every predicate is expressible as the sign of an angle polynomial
$f$ evaluated at a zero $t$ of an angle polynomial $g$.  A predicate is
degenerate (the sign is zero) when $t$ is a zero of a common factor of $f$ and
$g$.  We present an efficient degeneracy detection algorithm based on a one-time
factoring of every possible angle polynomial.  Our algorithm is 3500 times
faster than the standard algorithm based on greatest common divisor computation.
It reduces the share of degeneracy detection in our free space computations from
90\% to 0.5\% of the running time.
\end{abstract}

\section{Introduction}
\label{s-intro}

An implementation of a computational geometry algorithm is robust if for every
input the combinatorial output is correct and the numerical output is accurate.
The challenge is to implement the predicates in the algorithms: the signs of
algebraic expressions whose variables are input parameters.  A predicate is
\emph{degenerate} if its sign is zero.  A nondegenerate predicate can usually be
evaluated quickly, using machine arithmetic.  However, detecting that a
predicate is degenerate requires more costly computation.

\begin{figure}[htbp]
  \centering
  \begin{tabular}{ccc}
    \includegraphics[scale=0.8]{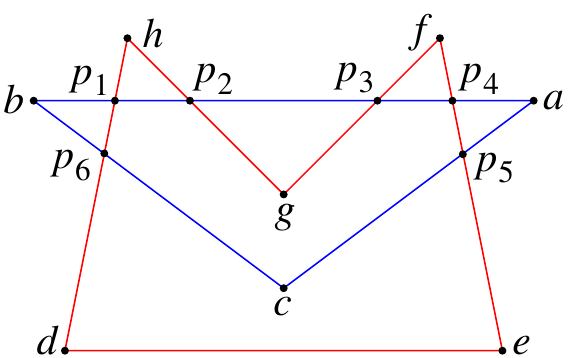} & 
    \includegraphics[scale=0.8]{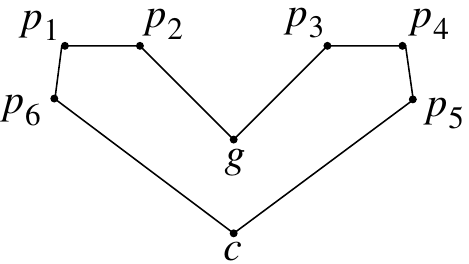} &
    \includegraphics[scale=0.8]{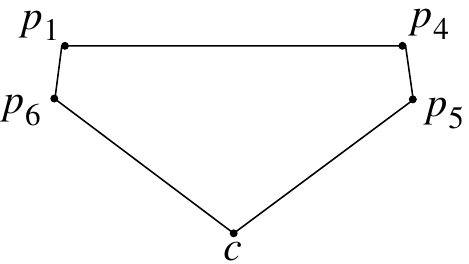}\\
    (a) & (b) & (c)
  \end{tabular}
  \caption{(a) Polygons, (b) intersection, and (c) convex hull.}
  \label{f-hull}
\end{figure}

We present research in degeneracy detection.  Prior research mainly addresses
degeneracy due to nongeneric input, such as the signed area of a triangle with
collinear vertices.  Such degeneracy is easily eliminated by input perturbation
\cite{halperin10}.  We address predicates, which we call \emph{identities,} that
are degenerate for all choices of the input parameters.  One example is the
signed area of a triangle $pab$ with $p$ the intersection point of line segments
$ab$ and $cd$, which is identical to zero when $p$ is replaced by its definition
in terms of the input.  This identity occurs when constructing the convex hull
of the intersection of two polygons.  Figure~\ref{f-hull} shows generic polygons
$abc$ and $defgh$ that intersect at points $\{p_1,\ldots,p_6\}$.  The convex
hull algorithm encounters an identity when it evaluates the signed area of any
three of $\{p_1,p_2,p_3,p_4\}$.  Triangulating the intersection of two polygons
involves similar identities.

Identities are common when the output of one algorithm is the input to another.
The second algorithm evaluates polynomials (the signed area in our example) on
arguments (the $p_i$ in our example) that are derived from input parameters by
the first algorithm.  When these algebraic expressions are rational, the
identities are amenable to polynomial identity detection \cite{schwartz80}.  We
are interested in identities involving more general algebraic expressions.

There are two general approaches to identity detection (Sec.~\ref{s-prior}).
One \cite{cgal} uses software integer arithmetic, computer algebra, and root
separation bounds to detect all degenerate predicates, including identities.
The second adds identity detection logic to computational geometry algorithms.
In the convex hull example, this logic checks if three points lie on a single
input segment.  The first approach can greatly increase the running time of the
software and the second approach can greatly increase the software development
time.

We present a new approach to identity detection that avoids the high running
time of numerical identity detection and the long development time of identity
detection logic.  The approach applies to a class of computational geometry
algorithms with a common set of predicates.  We write a program that enumerates
and analyzes the predicates.  The predicates are represented by algebraic
expressions with canonical variables.  The hull example requires 24 canonical
variables for the coordinates of the at most 12 input points that define the
three arguments of the signed area.  When implementing the algorithms, we match
their arguments against the canonical variables and use the stored analysis to
detect identities.

We apply this approach in constructing the free space of a polyhedron $R$ that
rotates around a fixed axis and translates freely relative to a polyhedron $O$
(Secs.~\ref{s-cs} and~\ref{s-sweep}).  For example, $R$ models a drone
helicopter and $O$ models a warehouse.  The configuration of $R$ is specified by
its position and its rotation angle.  The configuration space is the set of
configurations.  The free space is the subset of the configuration space where
the interiors of $R$ and $O$ are disjoint.  Robust and efficient free space
construction software would advance motion planning, part layout, assembly
planning, and mechanical design.

The structure of the free space is determined by the configurations where $R$
has four contacts with $O$.  A contact is determined by a vertex of $O$ and a
facet of $R$, a facet of $O$ and a vertex of $R$, or an edge of $O$ and an edge
of $R$.  The angle (under a rational parameterization) of a configuration with
four contacts is a zero of a univariate polynomial of degree~6, which we call an
\emph{angle polynomial}, whose coefficients are polynomials in the 48
coordinates of the 16 vertices of the 4 contacts.  Every predicate in free space
construction is expressible as the sign of an angle polynomial $f$ evaluated at
a zero $t$ of an angle polynomial $g$ (Sec.~\ref{s-identity}).  A predicate is
degenerate when $t$ is a zero of a common factor $h$ of $f$ and $g$.  It is an
identity when $h$ corresponds to a common factor of $f$ and $g$ considered as
multivariate polynomials in $t$ and in the vertex coordinates.

Neither prior identity detection approach is practical.  Detecting an identity
as a zero of the greatest common divisor of $f$ and $g$ is slow
(Sec.~\ref{s-results}).  Devising identity detection logic for every predicate
is infeasible because there are over 450,000,000 predicates and 13,000
identities (Sec.~\ref{s-analysis}).  We present an efficient identity detection
algorithm (Sec.~\ref{s-factoring}) based on a one-time analysis of the angle
polynomials (Sec.~\ref{s-offline}).

We enumerate the angle polynomials using canonical variables for the vertex
coordinates.  Working in the monomial basis is impractical because many of the
angle polynomials have over 100,000 terms.  Instead, we represent angle
polynomials with a data structure, called an \emph{a-poly}, that is a list of
sets of vertices (Sec.~\ref{s-apoly}).  The enumeration yields 1,000,000
a-polys, which we reduce to the 30,000 representatives of an equivalence
relation that respects factorization.  We construct a table of factors for the
equivalence class representatives in one CPU-day on a workstation.

We factor an angle polynomial by looking up the factors of its representative in
the table and substituting its vertex coordinates for the canonical variables.
We use the factoring algorithm to associate each zero of an angle polynomial
$g$ with an irreducible factor $h$.  Before evaluating a predicate at $t$, we factor its
angle polynomial $f$.  The predicate is an identity if $h$ is one of the factors.
Our algorithm is 3500 times faster than
computing greatest common divisors.  It reduces the share of degeneracy
detection in our free space computations from 90\% to 0.5\% of the running time
(Sec.~\ref{s-results}).  Sec.~\ref{f-discuss} provides guidelines for applying
table-based identity detection to other domains.

\section{Prior work}\label{s-prior}

Identity detection is the computational bottleneck in prior work by Hachenberger
\cite{hachenberger09} on computing the Minkowski sum of two polyhedra.  He
partitions the polyhedra into convex components and forms the union of the
Minkowski sums of the components.  Neighboring components share common,
collinear, or coplanar features, resulting in many identities in the union
operations.  Detecting the identities via the numerical approach (using CGAL)
dominates the running time.

Mayer et al \cite{mayer11} partially compute the free space of a convex
polyhedron that rotates around a fixed axis and translates freely relative to a
convex obstacle.  They report no identity detection problems.  Identities can be
detected using one rule: all polynomials generated from a facet of one
polyhedron and an edge of the other are the same up to sign and hence their
zeros are identical.  These polynomials correspond to our type I predicates for
general polyhedra (Sec.~\ref{s-sweep}).

We address identities in four prior works.  We \cite{kyung-sacks-milenkovic15}
compute polyhedral Minkowski sums orders of magnitude faster than Hachenberger
by using a convolution algorithm, which has fewer identities, and by detecting
identities with special case logic.  We \cite{sacks-milenkovic13} compute free
spaces of planar parts bounded by circular arcs and line segments.  The number
of identities is small, but the proof of completeness is lengthy.  We
\cite{sacks-butt-milenkovic17} compute free spaces of polyhedra where $R$
translates in the $xy$ plane and rotates around the $z$ axis.  The identity
detection logic is retrospectively confirmed using our new approach.  There are
816 equivalence classes with 290 in the basis versus 30,564 and 15,306 for the
4D configuration space (Sec.~\ref{s-offline}).

Finally, we \cite{sacks-milenkovic13b} find placements for three polyhedra that
translate in a box.  The algorithm performs a sequence of ten Minkowski sums and
Boolean operations, resulting in many identities.  One implementation handles
the identities as special cases.  A second implementation prevents identities
with a polyhedron approximation algorithm that rounds and perturbs the output of
each step.  The former is twice as fast as the latter and is exact, but took
months to develop and lacks a completeness proof.

\section{Free space}\label{s-cs}

This section begins our treatment of free space construction.  The polyhedra $R$
and $O$ have triangular facets.  Without loss of generality, we use the $z$ axis
as the axis of rotation.  We represent the rotation angle using a rational
parameterization of the unit circle.  A configuration $c$ of $R$ is a rotation
parameter $t$ and a translation vector $d$, denoted $c=(t, d)$.  It maps a point
$p$ to the point $c(p)=d+\Theta(t)p$ with
\begin{equation}\label{e-rat}
  \Theta(t)p=
  \left(\frac{(1-t^2)p_x-2tp_y}{1+t^2}, \frac{2tp_x+(1-t^2)p_y}{1+t^2}, p_z\right).
\end{equation}
A point set $P$ maps to $c(P)=\{c(p)\mid p\in P\}$.  The free space is
$\{c\mid O\cap c(R)=\emptyset\}$.

The boundary of the free space consists of \emph{contact configurations} $c$ at
which the boundaries of $c(R)$ and $O$ intersect but not the interiors.  The
generic contacts are a vertex $r_k$ of $R$ on a facet $o_ho_io_j$ of $O$, a
vertex $o_h$ of $O$ on a facet $r_ir_jr_k$ of $R$, and an edge $o_ho_i$ of $O$
tangent to an edge $r_jr_k$ of $R$.  The boundary has faces of dimension $k=0$
through $k=3$.  A face of dimension $k$ consists of configurations where $4-k$
contacts occur.

A necessary condition for contact is that the four vertices of the two features
are coplanar, so their tetrahedron has zero volume.  We substitute the
vertices---applying $c$ to those from $R$---into the volume formula
$(q-p)\times(u-p)\cdot(v-p)/6$ to obtain a \emph{contact expression}.  We
substitute Eq.~(\ref{e-rat}), multiply by 6, and clear the denominator of
$1+t^2$ to obtain a \emph{contact polynomial}, denoted $o_ho_io_j-r_k$,
$o_h-r_ir_jr_k$, or $o_ho_i-r_jr_k$ (Table~\ref{t-contact}).

\begin{table}[htbp]
  \caption{Contact polynomials.}\label{t-contact}
  \begin{tabular}{lll}
    denotation & contact expression & with\\
    $o_ho_io_j-r_k$ & $d\cdot u + u\cdot\Theta(t)r_k- u\cdot o_j$ &
    $u=(o_i-o_j)\times(o_h-o_j)$\\
    $o_h-r_ir_jr_k$ & $d\cdot\Theta(t)u + u\cdot r_k - o_h\cdot\Theta(t)u$ &
    $u=(r_i-r_k)\times(r_j-r_k)$\\
    $o_ho_i-r_jr_k$ & $d\cdot(u\times\Theta(t)v) + u\cdot\Theta(t)w + (u\times
    o_i)\cdot\Theta(t)v$
    & $u=o_h-o_i$, $v=r_j-r_k$, $w=r_j\times r_k$
  \end{tabular}
\end{table}

Computing the common zeros of four contact polynomials is a core task in free
space construction.  The polynomials have the form
$k_{i1}d_x+k_{i2}d_y+k_{i3}d_z+k_{i4}=0$ where the $k_{ij}$ are polynomials in
$t$.  They have a common zero at $t=t_0$ if the determinant $|k_{ij}|$ is zero
and the matrix $[k_{ij}]$ has a nonzero 3-by-3 left minor.  We construct the
faces of the free space boundary with a sweep algorithm whose events are zeros
of these determinants (Sec.~\ref{s-sweep}).  Moreover, the vertices are common
zeros of contact polynomials, as we explain next.

\begin{figure}[tbp]
\centerline{\includegraphics{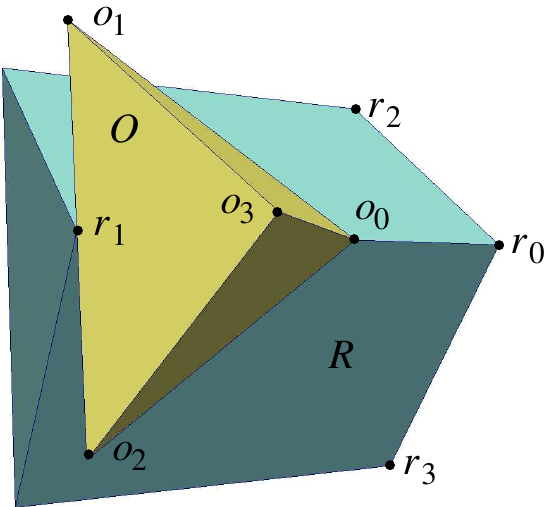}}
\caption{Identity in free space vertex predicate: $r_0$ is in the plane of
  $o_0o_1o_2$.}\label{f-intro}
\end{figure}

Figure~\ref{f-intro} depicts a zero of the contact polynomials
$p_1=o_0o_1o_2-r_1$, $p_2=o_1o_2o_3-r_1$, $p_3=o_0-r_0r_1r_2$, and
$p_4=o_0-r_0r_1r_3$.  For this to be a vertex, $c(r_0)c(r_2)$ cannot pierce
$o_0o_1o_2$ or else the interiors of $O$ and $c(R)$ would intersect.  The
edge/facet piercing test uses the signs of five contact polynomials, including
$o_0o_1o_2-r_0$.  The $p_1$ and $p_2$ contacts imply that $c(r_1)$ is on the
line of $o_1o_2$.  The $p_3$ and $p_4$ contacts imply that $o_0$ is on the line
of $c(r_0)c(r_1)$.  Since the line of $c(r_0)c(r_1)$ shares two points with the
plane of $o_0o_1o_2$, they are coplanar, $c(r_0)$ is in this plane, and so
$o_0o_1o_2-r_0$ is identically zero.  This identity resembles the signed area
identities (Sec.~\ref{s-intro}) in that a polynomial is evaluated on arguments
that are derived from the input.  However, we cannot apply
polynomial identity detection~\cite{schwartz80} because the arguments are not
rational functions of the input but rather the zeros of polynomials whose
coefficients are rational functions of the input.

\section{Predicates}\label{s-identity}

An \emph{angle polynomial} is a polynomial in $t$ that is used in free space
construction.  We show that every predicate is expressible as the sign of an
angle polynomial $f$ evaluated at a zero of an angle polynomial $g$.  The only
exception is the sign of a contact polynomial $p$ evaluated at a common zero $c$
of contact polynomials $\{p_1,p_2,p_3,p_4\}$.  We express this form like the
other predicates by constructing a polynomial $f$ such that $p(c)=0$ iff
$f(t_0)=0$ as follows.

Let $f$ be the determinant of $\{p_i,p_j,p_k,p\}$, with $\{p_i,p_j,p_k\}$,
$1\leq i < j < k\leq 4$, having a non-zero left 3-by-3 minor at $t=t_0$.  If
$f(t_0)=0$, $\{p_i,p_j,p_k,p\}$ are linearly dependent in $d$ at $t=t_0$, so $p$
is a linear combination of $\{p_i,p_j,p_k\}$ and $p(c)=0$.  If $p(c)=0$,
$\{p_i,p_j,p_k,p\}$ must be linearly dependent at $t=t_0$ because
$p_i(c)=p_j(c)=p_k(c)=0$ and hence $f(t_0)=0$.

A degenerate predicate has a polynomial $f$ such that $f(t_0)=0$ for $t=t_0$ a
zero of $g$.  In other words, $f$ and $g$ have a common factor $h$ and
$h(t_0)=0$.  This degeneracy is an identity if $h$ results from a common factor
of $f$ and $g$ considered as multivariate polynomials in $t$ and in the
canonical vertex coordinates.  We make such common factor detection fast by
enumerating the canonical polynomials, factoring them, and storing the results
in a table.

\section{Angle polynomials}\label{s-apoly}

We represent an angle polynomial with an \emph{a-poly:} a list of elements of
the form $L_O-L_R$ with $L_O$ and $L_R$ lists of vertices of $O$ and of $R$ in
increasing index order.  Elements are in order of increasing $|L_O|+|L_R|$, then
increasing $|L_O|$, then increasing vertex index (lexicographically).
There are three kinds of a-polys.

The first kind represents the angle polynomials at whose zeros four contacts can
occur.  It has three types of elements.  A 1-contact denotes a contact
polynomial.  A 2-contact $o_h-r_ir_j$ denotes 1-contacts $o_h-r_ir_jr_k$ and
$o_h-r_ir_jr_l$ that jointly constrain $o_h$ to the line $r_ir_j$; likewise
$o_ho_i-r_j$.  A 3-contact $o_h-r_i$ denotes three polynomials whose zeros are
the configurations where the two vertices coincide.  A list of elements that
together denote four polynomials comprises an a-poly.

The second kind represents an angle polynomial that is zero when an edge of one
polyhedron and two edges of the other polyhedron are parallel to a common plane.
It has three elements of types $o_io_j-$ and $-r_ir_j$, for example
$(o_1o_2-,-r_1r_2,-r_4r_5)$.  However, if the two edges share a vertex, we
contract $(o_ho_i-,-r_jr_k,-r_jr_l)$ to $(o_ho_i-r_jr_kr_l)$, corresponding to
an edge parallel to a facet.  Likewise, $(o_ho_io_j-r_kr_l)$.  The third kind
corresponds to the 3-by-3 left minor (Sec.~\ref{s-cs}): the $d$-coefficients of
three contact polynomials.  The $d$-coefficients of $o_io_jo_k-r_l$ and
$o_i-r_ir_jr_k$ do not depend on $r_l$ and $o_i$, hence the elements are of type
$o_io_jo_k-$, $-r_ir_jr_k$, or $o_ho_i-r_jr_k$.

The derivation of the angle polynomials from their a-polys is as follows.

\subparagraph*{Four 1-contacts} The contact expressions (Table~\ref{t-contact})
have the form $d\cdot n+k$.  The vectors $n$ have the form $a$,
$\Theta(t)a$, or $a\times\Theta(t)b$ and the summands of the scalar
$k$ have the form $a\cdot b$ or $a\cdot\Theta(t)b$ with $a$ and $b$
constant vectors.  The angle polynomial is the numerator of
\begin{displaymath}\label{e-con}
\left|\begin{array}{cccc}
n_{1x} & n_{1y} & n_{1z} & k_1\\
n_{2x} & n_{2y} & n_{2z} & k_2\\
n_{3x} & n_{3y} & n_{3z} & k_3\\
n_{4x} & n_{4y} & n_{4z} & k_4
\end{array}\right|
.
\end{displaymath}
Expanding in terms of minors using the last column, $k_1$ has minor
$n_2\cdot(n_3\times n_4)$.  Using the formula, $u\times(v\times w)= (u \cdot w)
v - (u \cdot v) w$, each minor reduces to a sum of terms of the form $a\cdot b$,
$a\cdot\Theta(t)b$, or $(a\cdot\Theta(t)b)(c\cdot\Theta(t)d)$.  Applying
equation~(\ref{e-rat}) and clearing the denominator reduces $k_1$, $k_2$, and
$k_3$ to quadratics in $t$ and reduces the minors to quartics in $t$, so the
angle polynomial has degree~6.

\subparagraph*{One 2-contact and two 1-contacts} For a 2-contact $o_io_j-r_k$,
$c(r_k)=d+\Theta(t)r_k$ is on the line of $o_io_j$, so $d$ is on the line
through $o_i-\Theta(t) r_k$ and $o_j-\Theta(t) r_k$.  We intersect this line
with the planes of the other two contact polynomials.  We express the line as
$\lambda u+v$ with $u=o_j-o_i$ and $v=o_j-\Theta(t)r_k$, compute the values
$\lambda_i=-(n_i\cdot v+k_i)/(n_i\cdot u)$ where the line intersects the two
planes $n_i\cdot p+k_i$, set $\lambda_1=\lambda_2$, and cross multiply to obtain
a quartic angle polynomial.  Similarly 2-contact $o_i-r_jr_k$ corresponds to a
line with $u=\Theta(t)(r_j-r_k)$ and $v=o_i-\Theta(t)r_j$.

\subparagraph*{Two 2-contacts} The expression is the signed volume of the four points
that define the lines of the 2-contacts, which yields a quartic angle
polynomial.  Figure~\ref{f-intro} illustrates this situation.

\subparagraph*{One 3-contact and one 1-contact} For a 3-contact $o_i-r_j$,
$o_i=d+\Theta(t)r_j$. We substitute $d=o_i-\Theta(t)r_j$ into the 1-contact
polynomial to obtain a quadratic angle polynomial.

\subparagraph*{Other kinds}
The second, $(-r_{i_1}r_{j_1},-r_{i_2}r_{j_2}, o_{i_3}o_{j_3}-)$ and
$(-r_{i_1}r_{j_1}, o_{i_2}o_{j_2}-,o_{i_3}o_{j_3}-)$, has expressions that yield
quadratic angle polynomials
$\Theta(t)((r_{j_1}-r_{i_1})\times(r_{j_2}-r_{i_2}))\cdot(o_{j_3}-o_{i_3})$ and
$\Theta(t)(r_{j_1}-r_{i_1})\cdot((o_{j_2}-o_{i_2})\times(o_{j_3}-o_{i_3}))$.
The second has expression $(n_1\times n_2)\cdot n_3$, where $o_io_jo_k-$ has
normal $n=(o_j-o_i)\times(o_k-o_i)$, $-r_ir_jr_k$ has normal
$n=\Theta(t)((r_j-r_i)\times(r_k-r_i))$, and $o_ho_i-r_jr_k$ has normal
$n=(o_i-o_h)\times\Theta(t)(r_k-r_j)$.  The third has the same polynomials as
the quartic minor polynomials above in the four 1-contacts case.

\section{Factoring}\label{s-factoring}

This section gives an algorithm for factoring angle polynomials represented as
a-polys.  Two a-polys are \emph{equivalent} if a vertex bijection maps one to
the other.  The bijection maps a factorization of one to a factorization of the
other.  The factoring algorithm uses a table that contains the factorization of
a representative a-poly from each equivalence class.  Sec.~\ref{s-offline}
explains how we constructed this table.  It is available in the web directory
\href{http://www.cs.miami.edu/~vjm/robust/identity}{\url{http://www.cs.miami.edu/~vjm/robust/identity}}.

Identity detection requires that an irreducible polynomial be denoted by a
unique a-poly, so one can detect if different a-polys have a common
factor.  One problem is that nonequivalent a-polys can denote the same
polynomial.  We solve this problem by selecting the factors in the table from a
minimal subset of the equivalence classes that we call \emph{basis} classes.  A
second problem is that equivalent a-polys can denote the same polynomial.
This problem is so rare that we can record all the basis a-polys that
generate a factor, called its \emph{factor set,} and select a unique one during
factoring.  The factoring algorithm maps an input a-poly to the
representative of its equivalence class, obtains the factor sets of the
representative from the table, and applies the inverse map to obtain sets of
a-polys in the variables of the input a-poly.  To achieve uniqueness, it
selects the lexicographical minimum from each set, using a vertex order that we
indicate by the $o$ and $r$ indices.

\subsection{Mapping an a-poly to its representative}

The mapping algorithm generates the permutations of the input a-poly such
that the elements remain increasing in $|L_O|+|L_R|$ then in $|L_O|$ (but
disregarding the lexicographical order of vertex indices).  For each permuted
contact list, it assigns each vertex an indicator: a bit string in which a 1 in
position $k$ indicates that the vertex appears in the $k$th element.  It labels
a permutation with its $O$ vertex indicators in decreasing order followed by its
$R$ vertex indicators in decreasing order.  It selects the permutation with the
largest label, replaces the $i$th $O$ vertex in indicator order by the canonical
vertex $o_i$, and likewise for the $R$ vertices and $r_i$.  If two vertices have
the same indicator, both orders yield the same output because the indices of
an a-poly are placed in increasing order.

For example, in a-poly $(o_{27}-r_{22}r_{66}r_{86}, o_{43}-r_{22}r_{66}r_{86},
o_{27}o_{51}-r_{15}r_{86},o_{27}o_{43}-r_{75}r_{86})$, $o_{27}$ has indicator
$1011$, and the indicator list is 1011,0101,0010;1111,1100,1100,0010,0001.  The
permutations swap the first two and/or last two elements.  Swapping the last two
yields the maximal indicator list 1011,0110,0001;1111,1100,1100,0010,0001 and
the representative $(o_0-r_0r_1r_2,o_1-r_0r_1r_2,o_0o_1-r_0r_3,o_0o_2-r_0r_4)$.
In the table, this
representative has factor sets $\{(o_0o_1-r_0r_1r_2)\}$ and
$\{(-o_0o_1o_2, o_0o_1-r_0r_3, o_0o_2-r_0r_4)\}$.  The inverse vertex mapping
yields the factors $(o_{27}o_{43}-r_{22}r_{66}r_{86})$ and
$(-r_{22}r_{66}r_{86}, o_{27}o_{43}-r_{75}r_{86}, o_{27}o_{51}-r_{15}r_{86})$.

\subsection{Uniqueness}\label{s-uniqueness}

\begin{figure}[htbp]
\begin{tabular}{cc}
\includegraphics[height=1.4in]{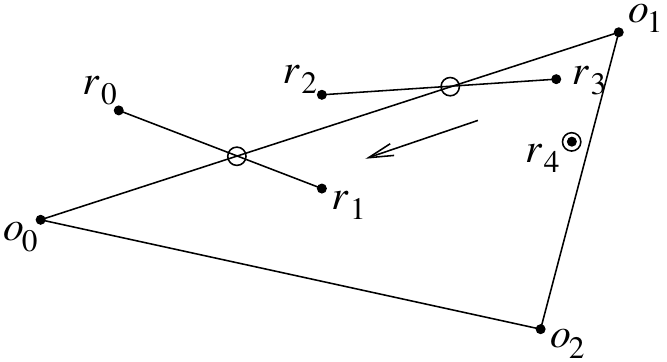} &
\includegraphics[height=1.4in]{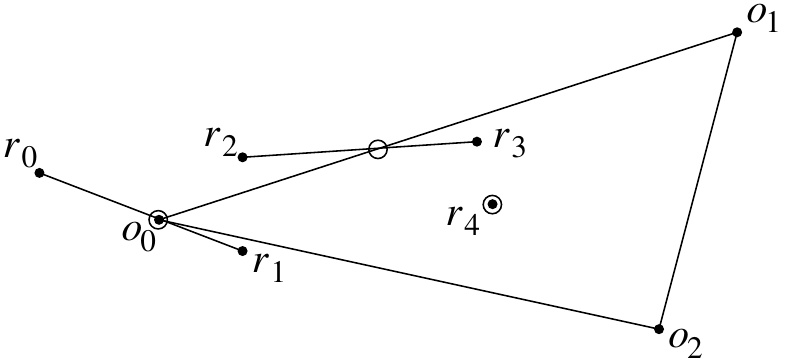}\\
(a) & (b)
\end{tabular}
\caption{Equivalent a-polys: (a) translation parallel to $o_0o_1$
  preserves circled contacts $o_0o_1-r_0r_1$, $o_0o_1-r_2r_3$, and
  $o_0o_1o_2-r_4$; (b) a-poly $(o_0-r_0r_1, o_0o_1-r_2r_3, o_0o_1o_2-r_4)$.}
\label{f-cat3}
\end{figure}

Fig.~\ref{f-cat3} illustrates how a-polys can have the same polynomial.  The
1-contacts $o_0o_1-r_0r_1$, $o_0o_1-r_2r_3$, and $o_0o_1o_2-r_4$ determine the
angle parameter $t$ because they are invariant under translation parallel to
$o_0o_1$.  Translating $R$ parallel to $o_0o_1$ until one element becomes a
2-contact yields an a-poly $C_i$ that is zero at the same
$t$ values.  If $r_0r_1$ hits $o_0$, $C_1=(o_0-r_0r_1, o_0o_1-r_2r_3,
o_0o_1o_2-r_4)$ (Fig.~\ref{f-cat3}b) and if $r_2r_3$ hits $o_1$,
$C_2=(o_1-r_2r_3, o_0o_1-r_0r_1, o_0o_1o_2-r_4)$.  These a-polys are
equivalent under the map from $C_1$ to $C_2$: $o_0\rightarrow o_1$,
$o_1\rightarrow o_0$, $o_2\rightarrow o_2$, $r_0\rightarrow r_2$,
$r_1\rightarrow r_3$, $r_2\rightarrow r_0$, $r_3\rightarrow r_1$,
$r_4\rightarrow r_4$.  There are also $C_3$ and $C_4$ where $r_0r_1$ hits $o_1$
or $r_2r_3$ hits $o_0$ and $C_5=(o_0o_2-r_4,o_0o_1-r_0r_1, o_0o_1-r_2r_3)$ and
$C_6=(o_1o_2-r_4,o_0o_1-r_0r_1, o_0o_1-r_2r_3)$ where $r_4$ hits $o_0o_2$ or
$o_1o_2$.  $C_5$ and $C_6$ are not equivalent to $C_1$, $C_2$, $C_3$, and $C_4$
because their first element has two $O$ vertices, not one.

The a-poly
$(o_1-r_0r_1r_2,o_0o_1-r_3r_4,o_0o_1-r_5r_6,o_oo_1o_2-r_7)$ maps to the
representative $(o_0-r_0r_1r_2,o_0o_1-r_3r_4,o_0o_1-r_5r_6,o_oo_1o_2-r_7)$ with
factor sets $\{(o_0o_1-r_0r_1r_2)\}$ and
\begin{eqnarray*}
\{(o_0-r_3r_4,o_0o_1-r_5r_6,o_0o_1o_2-r_7),
(o_0-r_5r_6,o_0o_1-r_3r_4,o_0o_1o_2-r_7),\\
(o_1-r_3r_4,o_0o_1-r_5r_6,o_0o_1o_2-r_7),
(o_1-r_5r_6,o_0o_1-r_3r_4,o_0o_1o_2-r_7)\}
\end{eqnarray*}
because the second factor is equivalent to $C_1,\ldots,C_4$.  Its factor set
does not contain a-polys equivalent to $C_5$ and $C_6$ because their class is
not in the basis.  The inverse map (in this case swapping $o_0$ and $o_1$)
results in factor sets $\{(o_0o_1-r_0r_1r_2)\}$ and
\begin{eqnarray*}
\{(o_1-r_3r_4,o_0o_1-r_5r_6,o_0o_1o_2-r_7),
(o_1-r_5r_6,o_0o_1-r_3r_4,o_0o_1o_2-r_7),\\
(o_0-r_3r_4,o_0o_1-r_5r_6,o_0o_1o_2-r_7),
(o_0-r_5r_6,o_0o_1-r_3r_4,o_0o_1o_2-r_7) \}.
\end{eqnarray*}
The algorithm selects the (minimal) third element
$(o_0-r_3r_4,o_0o_1-r_5r_6,o_0o_1o_2-r_7)$ of the second set as the second
factor.

\section{Constructing the table of factors}\label{s-offline}

This section explains how we enumerate the a-poly equivalence classes
(Sec.~\ref{s-rep}), factor the class representatives and select a basis
(Sec.~\ref{s-basis}), and construct the table of factors (Sec.~\ref{s-table}).
The factoring algorithm is probabilistic and depends on the completeness
assumption that the factors of a-polys are a-polys.  If this
assumption were false, the algorithm would have failed.  We verify the table to
a high degree of certainty using standard techniques for testing polynomial
identities (Sec.~\ref{s-analysis}).

\subsection{Equivalence classes}
\label{s-rep}

We enumerate the a-poly classes as follows.  Let $a_i$, $b_i$, $c_i$, $d_i$,
$e_i$, and $f_i$ denote vertices and define
$s_i=\{a_i-d_i\}$,
$t_i=\{a_i-d_ie_i,a_ib_i-d_i\}$,
$u_i=\{a_i-d_ie_if_i,a_ib_i-d_ie_i,a_ib_ic_i-d_i\}$,
$v_i=\{a_ib_i-c_id_ie_i,a_ib_ic_i-d_ie_i\}$,
$w_i=\{-a_ib_i,a_ib_i-\}$,
$x_i=\{-a_ib_ic_i,a_ib_i-c_id_i,a_ib_ic_i-\}$.
We generate the a-polys that are lists of $k$-contacts
with the sets
$s_1\times{}u_2$ (a 3-contact and a 1-contact), $t_1\times{}t_2$ (two
2-contacts), $t_1\times{}u_2\times{}u_3$ (a 2-contact and two 1-contacts), and
$u_1\times{}u_2\times{}u_3\times{}u_4$ (four 1-contacts).  We generate the other
kinds of a-polys with the sets $v_1$ (edge parallel to facet), $w_1\times{}w_2\times{}w_3$ (edges parallel to plane), and
$x_1\times{}x_2\times{}x_3$ (3-by-3 left minor).  For each element of each set, we assign $O$
vertices to the $a_i,b_i,c_i$ in every possible manner.  Starting from $o_0$, we
assign increasing indices to the vertices of an
edge or a facet.  We assign $R$ vertices to $d_i,e_i,f_i$ likewise.  The highest
index is 11.

For example, $s_1\times{}u_2=
\{(a_1-d_1,a_2-d_2e_2f_2),(a_1-d_1,a_2b_2-d_2e_2),(a_1-d_1,a_2b_2c_2-d_2)\}$.
We must set $a_1=o_0$.  We can set $a_2=o_0$ or $a_2=o_1$ because $a_2$ is in a
different feature, and then assign increasing indices to $b_2$ and $c_2$.
Similarly for $d_1$, $d_2$, $e_2$, and $f_2$.  The results are\\ 
$\{(o_0-r_0,o_0-r_0r_1r_2),
(o_0-r_0,o_1-r_0r_1r_2),
(o_0-r_0,o_0-r_1r_2r_3),
(o_0-r_0,o_1-r_1r_2r_3),$\\
\hspace*{1ex}$(o_0-r_0,o_0o_1-r_0r_1),
(o_0-r_0,o_1o_2-r_0r_1),
(o_0-r_0,o_0o_1-r_1r_2),
(o_0-r_0,o_1o_2-r_1r_2),$\\
\hspace*{1ex}$(o_0-r_0,o_0o_1o_2-r_0),
(o_0-r_0,o_1o_2o_3-r_0),
(o_0-r_0,o_0o_1o_2-r_1),
(o_0-r_0,o_1o_2o_3-r_1)\}$.

The enumeration yields about one million a-polys.  Generating their
representatives (Sec.~\ref{s-factoring}) and removing duplicates yields 30,564
equivalence class representatives.

\subsection{Basis classes}\label{s-basis}

We factor the representatives probabilistically.  We replace the canonical
coordinates of $o_0,\ldots,o_{11}$ and $r_0,\ldots,r_{11}$ with random integers,
construct the resulting univariate integer polynomials, and factor them with
Mathematica.  An irreducible univariate implies that the canonical polynomial
is irreducible; the converse is true with high probability.

An a-poly depends on a vertex if its univariate changes when the coordinates of
the vertex are assigned different random integers.  For example,
$(o_0o_1-r_0,o_2-r_1r_2r_3,o_2-r_1r_2r_4)$ does not depend on $r_3$ and $r_4$
because $o_2-r_1r_2r_3$ and $o_2-r_1r_2r_4$ can be replaced by $o_2-r_1r_2$:
$o_2$ in contact with $r_1r_2r_3$ and $r_1r_2r_4$ is equivalent to $o_2$ in
contact with $r_1r_2$.  An a-poly is \emph{complete} if it depends on all of its
vertices.  An angle polynomial is \emph{contiguous} if it depends on
$o_0,o_1,\ldots,o_{l}$ and $r_0,r_1,\ldots,r_{m}$ for some $l$ and $m$.  A
complete representative is also contiguous because we assign vertices to the
indicators contiguously (Sec.~\ref{s-factoring}).

We select a basis of complete, contiguous, and irreducible a-polys, represented
by the representatives of their classes.  (We prove that such a basis exists by
verifying the table of factors (Sec.~\ref{s-analysis}).)  We construct a map $I$
from the univariate of each basis a-poly to the set of basis a-polys that
generate it.  In the Sec.~\ref{s-uniqueness} example, $C_1$ is in the basis and
generates a univariate $p$.  The equivalent a-polys $C_2$, $C_3$, and $C_4$ also
generate $p$, so $I(p)=\{C_1,C_2,C_3,C_4\}$.  Although $C_5$ and $C_6$ also
generate $p$, they are not in $I(p)$ because they belong to another (necessarily
non-basis) equivalence class.

The algorithm visits each representative $\rho$.  If $\rho$ is complete and its
univariate $p$ is irreducible but $I(p)=\emptyset$, the algorithm adds $\rho$ to
the basis representatives, permutes its vertices in every way, calculates the
univariate $u$ for each resulting a-poly $a$, and adds $a$ to the set $I(u)$.

The condition $I(p)=\emptyset$ prevents adding an a-poly to the
basis whose univariate is already generated by a member of a basis class.  For
example, $\rho_1=(-r_0r_1r_2,-r_0r_3r_4,o_0o_1-r_0r_5)$ is assigned to the
basis.  Later, $\rho_2=(o_0-r_0r_1,o_0-r_2r_3r_4,o_0o_1-r_2r_5)$ is complete and
has an irreducible univariate that is generated by
$(-r_0r_1r_2,-r_2r_3r_4,o_0o_1-r_2r_5)$, which is a permutation of $\rho_1$.
Since every permutation of $\rho_1$ is in $I$, $\rho_2$ is not assigned to the
basis.

\subsection{Factor table}\label{s-table}

The factor table provides a list of factor sets for each representative.  For a
basis representative $\rho$ with univariate $p$, the list is $\langle
I(p)\rangle$.  If $\rho$ is not in the basis, we process each factor $f$ of $p$
as follows.

1) Determine which vertices $f$ depends on. Randomly change a vertex of $\rho$,
generate the new univariate, and factor it.  If $f$ is not a factor, it depends
on the vertex.

2) Rename the vertices of $\rho$ to obtain a $\rho'$ for which the
  factor $f'$ that corresponds to $f$ is contiguous.  Let $f$ depend on
$o_{i_0},o_{i_1},\ldots,o_{i_{m'}}$ and
$r_{j_0},r_{j_1},\ldots,r_{j_{n'}}$ but not on
$o_{i_{m'+1}},o_{i_{m'+2}},\ldots,o_{i_m}$ and
$r_{j_{n'+1}},r_{j_{n'+2}},\ldots,r_{j_n}$.  Substitute $o_{i_k}\rightarrow
o_k$ for $k=1,\ldots,m$ and $r_{i_k}\rightarrow r_k$ for $k=1,\ldots,n$.

3) Find the factor $f'$ of $\rho'$ that depends on $o_0,\ldots,o_{m'}$ and
$r_0,\ldots,r_{n'}$, and has the same degree as $f$.  (This factor is unique;
otherwise, we would consider every match.)

4) Look up $I(f')$ and invert the vertex substitution to obtain a factor
set.

For example, the univariate of
$\rho=(o_0o_1-r_0r_1,o_2o_3o_4-r_2,o_2o_3o_4-r_3,o_5o_6o_7-r_4)$ has a factor
$f$ that depends on all its variables and a quadratic factor $g$ that depends on
$o_2,o_3,o_4,r_2,r_3$.  The factor set of $f$ is
$I(f)=\{(o_2o_3o_4-,o_5o_6o_7-,o_0o_1-r_0r_1)\}$.  To obtain the factor set of
$g$, substitute ${o_2,o_3,o_4,o_0,o_1,r_2,r_3,r_0,r_1}\rightarrow{o_0,o_1,o_2,o_3,o_4,r_0,r_1,r_2,r_3}$.  The
quadratic factor $g'$ of
$\rho'=(o_3o_4-r_2r_3,o_0o_1o_2-r_0,o_0o_1o_2-r_1,o_5o_6o_7-r_4)$ depends on
$o_0o_1o_2$ and $r_0r_1$, and $I(g')=\{(o_0o_1o_2-r_0r_1)\}$.  Inverting the
vertex substitution yields the second factor set of $\rho$:
$\{(o_2o_3o_4-r_2r_3)\}$.

To save space, we do not add entries to $I$ corresponding to permutations of
basis representatives with degree-6 univariates because they cannot be proper
factors.  To test if $\rho$ with irreducible degree-6 univariate $p$ is in the
basis, we generate the permutations of $\rho$ and their univariates.  If none
has an entry in $I$, $\rho$ is in the basis, and we add an entry for $p$ to $I$.
If the univariate $p'$ of a permutation has an entry in $I$, the sole factor set
of $\rho$ is the result of applying the inverse of the permutation to $I(p')$.

\subsection{Analysis}\label{s-analysis}

Out of 30,564 representatives, 15,306 are basis, 991 are constant, 3840 are
irreducible but non-basis, 8263 have two factors (including 260 squares), and
2164 have three factors (including 6 cubes).  Since a predicate is an a-poly
evaluated on a zero of a basis polynomial, we listed $450,000,000\approx
30,564\cdot 15,306$ predicates in the introduction.  Likewise, we stated the
number of identities as $13,000\approx 1\cdot 3840 + 2\cdot 8263 + 3\cdot 2164$
ways of evaluating a non-basis polynomial on the zero of a factor.  Of the
irreducible representative polynomials, 363 have two basis a-polys, 50 have
three, 194 have four, and 194 have six.

Each factorization $f_1f_2\cdots f_m|f$ is equivalent to a polynomial identity
$f-a f_1f_2\cdots f_m=0$ for some constant $a$.  Instead of analyzing the
probability of failure of the algorithm, we verify the identities
probabilistically using Schwartz's lemma \cite{schwartz80}.  We use random
20-bit values modulo a prime $p$.  The first substitution determines $a$ and the
rest verify the identity.  Verifying all 15,258 factorizations once takes 2
seconds and 10 minutes of verification reduces the probability of an error to
below $10^{-1000}$.  This also constitutes a probabilistic proof of the
completeness assumption.

The running time for factor table construction was one CPU day.  All but one CPU
hour was spent generating the permutations of the degree-six irreducible
polynomials to test if they are in the basis.  The worst case is four contacts
between $O$ vertices and $R$ facets or vice versa, which have about 70 billion
permutations.  The tests all succeeded, so perhaps we could have avoided this
cost by proving a theorem.

\section{Contact set subdivision}\label{s-sweep}

We continue our discussion of free space with an algorithm for constructing the
faces of its boundary.  The \emph{contact set} of a vertex/facet or edge/edge
pair is the set of configurations where the two features are in contact.  We
construct the subdivision of a contact set induced by the other contact sets.
Each predicate used in the construction has the same zero set as an a-poly, so
identity detection applies.  The remaining (and substantial) step in free space
construction is to stitch the faces into a boundary representation of the free
cells.

\subparagraph*{Contact sets} The contact set for a triangle $o_ho_io_j$ of $O$ and a
vertex $r_k$ of $R$ is the set of configurations $c=(t,d)$ such that
$c(r_k)=d+\Theta(t)r_k$ lies on $o_ho_io_j$.  Hence, $d$ lies on the triangle
$o_h-\Theta(t) r_k, o_i-\Theta(t) r_k, o_j-\Theta(t) r_k$.  This triangle is a
cross-section of the contact set whose vertices are rational functions of $t$.
Similarly, the contact set for $o_h$ and $r_ir_jr_k$ is the parameterized
triangle $o_h-\Theta(t) r_i,o_h-\Theta(t) r_j,o_h-\Theta(t) r_k$, and the
contact set for $o_ho_i$ and $r_jr_k$, is the parameterized parallelogram
$o_h-\Theta(t) r_j,o_h-\Theta(t) r_k, o_i-\Theta(t) r_j, o_i-\Theta(t) r_k$.

\subparagraph*{Contact facets} We are only interested in the portion of a contact set
that is on the boundary of the free space.  A necessary condition is that the
interiors of $O$ and $c(R)$ are disjoint in a neighborhood of their point of
contact.  A \emph{contact facet} is the restriction of a contact set to the
intervals of $t$ values where this condition holds.  We express the condition in
terms of the signs of a-polys (ignoring the vertex index order rule),
hence the intervals are bounded by zeros of a-polys.  For
$o_ho_io_j$ and $r_k$, $(o_ho_io_j-r_kr_l)$ must be positive for every edge
$r_kr_l$; likewise for $o_h$ and $r_ir_jr_k$.  For $o_ho_i$ and $r_jr_k$, let
$o_ho_i$ be incident on the triangles $o_ho_im_1$ and $o_io_hm_2$, let $r_jr_k$
be incident on the triangles $r_jr_kn_1$ and $r_kr_jn_2$, and let $s_x$ and
$t_x$ be the signs of $(o_ho_im_x-r_jr_k)$ and $(o_ho_i-r_jr_kn_x)$ for $x=1,2$.
The interiors are locally disjoint if $s_1=s_2=-t_1=-t_2$.

The parameterized edges of a contact facet are called contact facet edges and
are in the zero sets of 2-contacts.  The parameterized vertices are called
contact facet vertices and are in the zero sets of 3-contacts.

\subparagraph*{Contact facet subdivision}
At a fixed $t$, each contact facet (triangle or parallelogram) is intersected
and subdivided by other contact facets.  The intersection of two facets is
called an FF-edge, and the intersection of a facet edge and a facet is called an
EF-vertex.  The endpoints of an FF-edges are facet vertices or EF-vertices.  Two
FF-edges intersect at an FFF-vertex, the intersection of three facets.

\begin{figure}[htbp]
\centering
\begin{tabular}{cc}
\includegraphics[scale=.5]{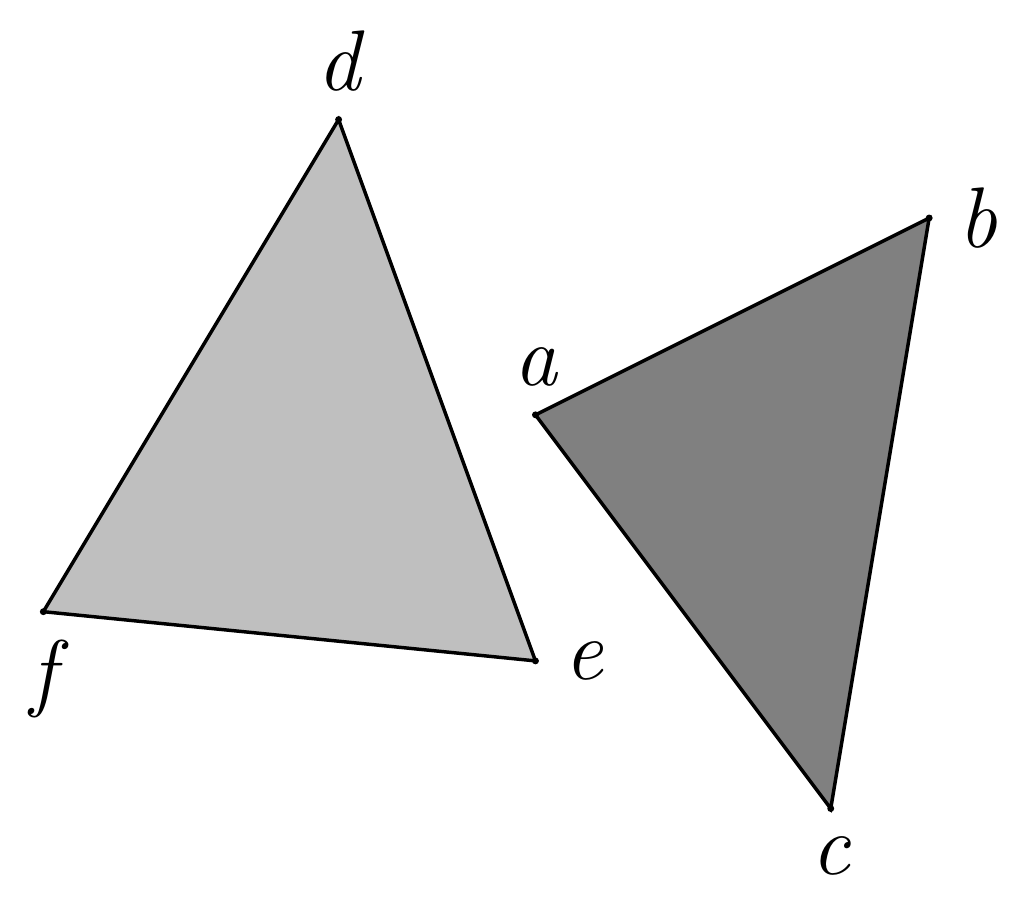} & 
\includegraphics[scale=.5]{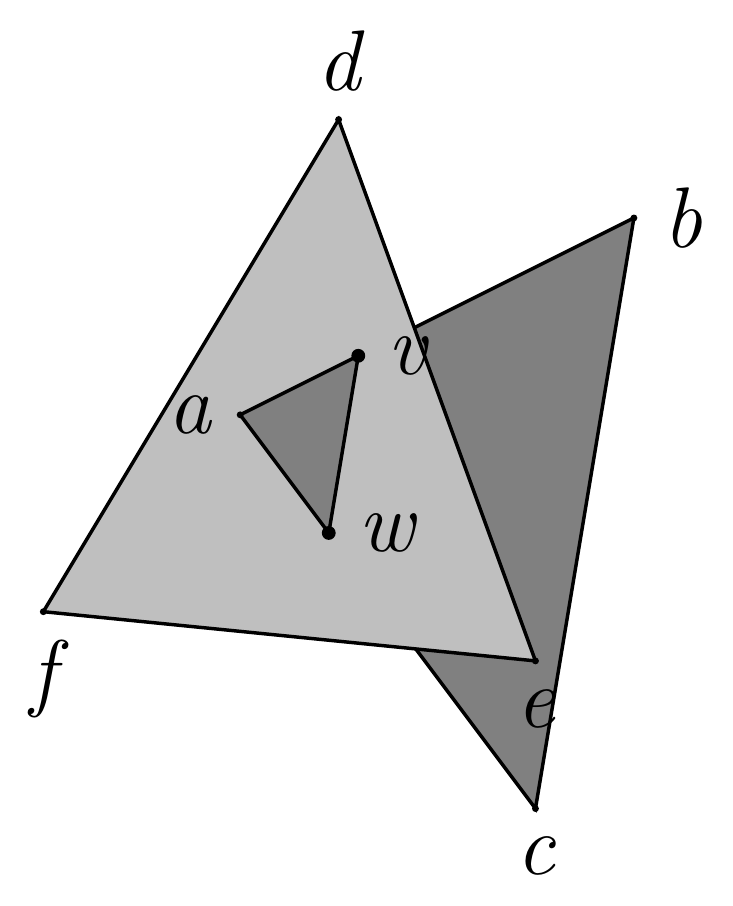} \\
\multicolumn{2}{c}{(a)} \\
\includegraphics[scale=.5]{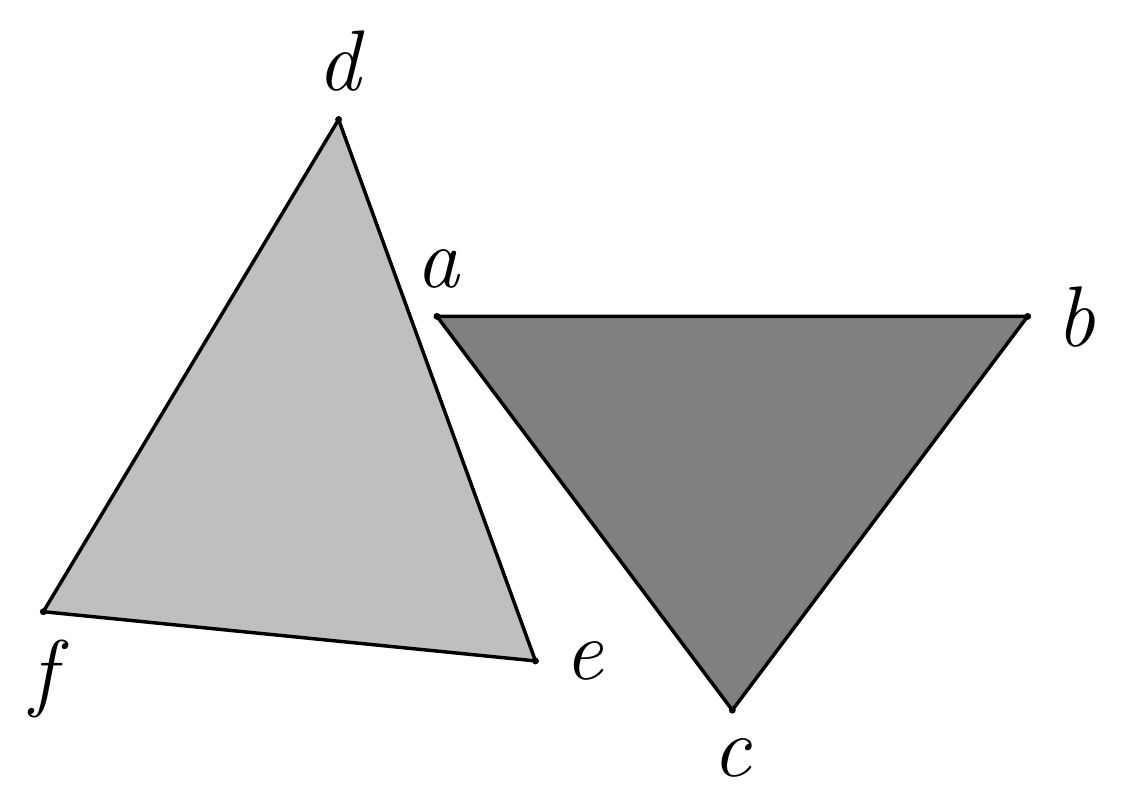} & 
\includegraphics[scale=.5]{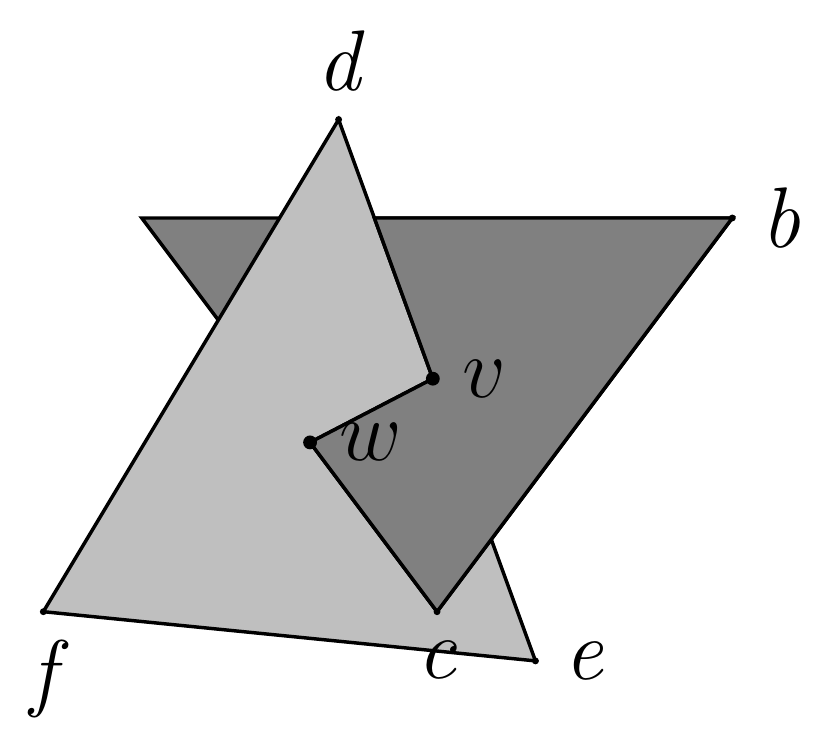}\\
\multicolumn{2}{c}{(b)}
\end{tabular}
\caption{Type~II structure changes: (a) vertex--facet, (b) edge--edge.}\label{f-2}
\end{figure}

\begin{figure}[htbp]
\centering
\begin{tabular}{cc}
\includegraphics[scale=.2]{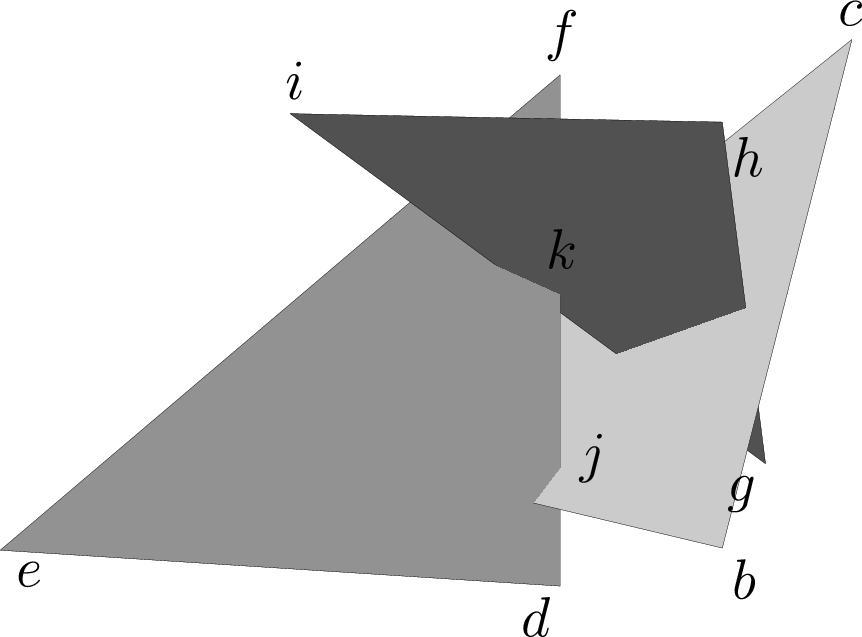} & 
\includegraphics[scale=.2]{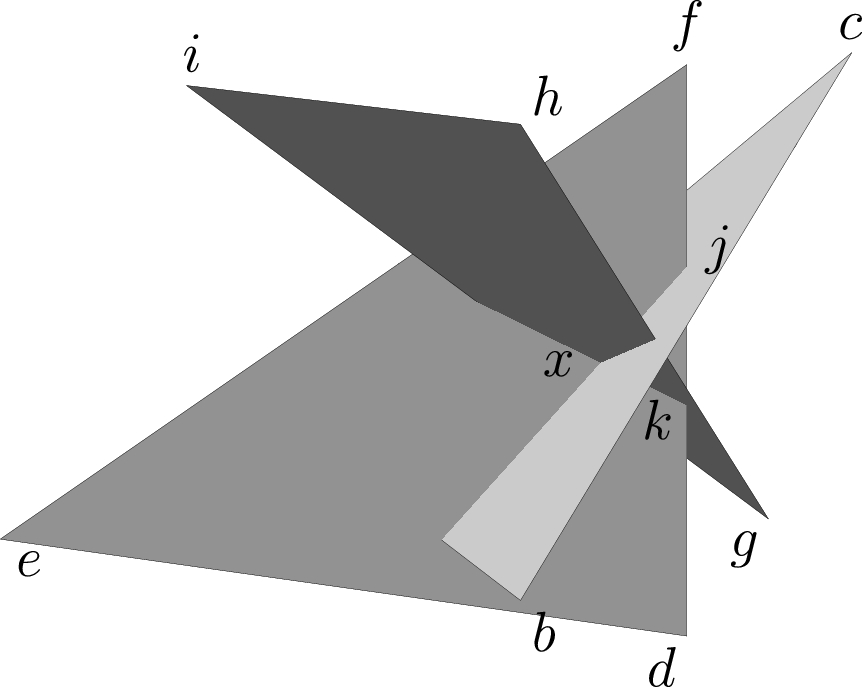}
\end{tabular}
\caption{Type~III structure change.}\label{f-3}
\end{figure}

\begin{figure}[htbp]
\centering
\begin{tabular}{cc}
\includegraphics[scale=.5]{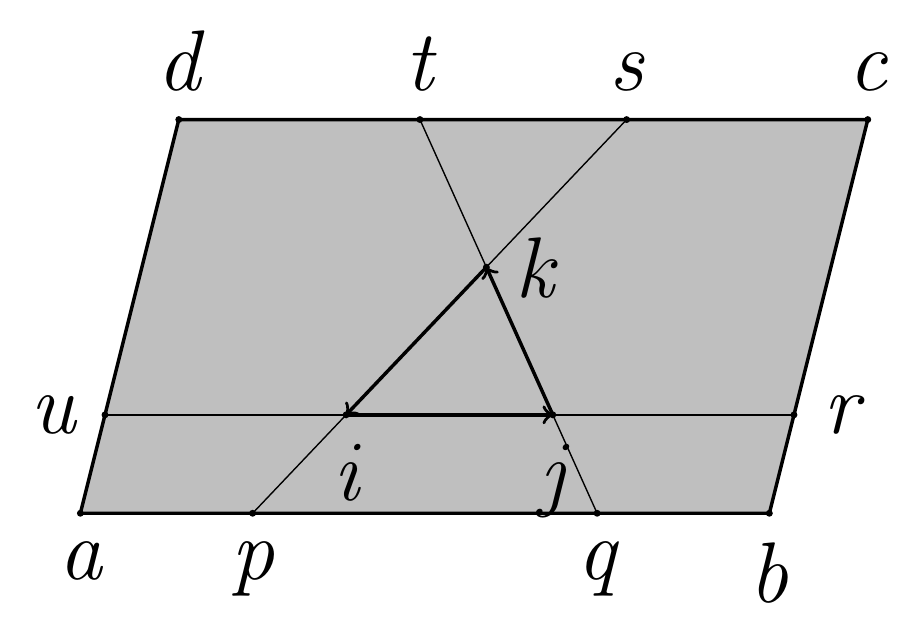} & 
\includegraphics[scale=.5]{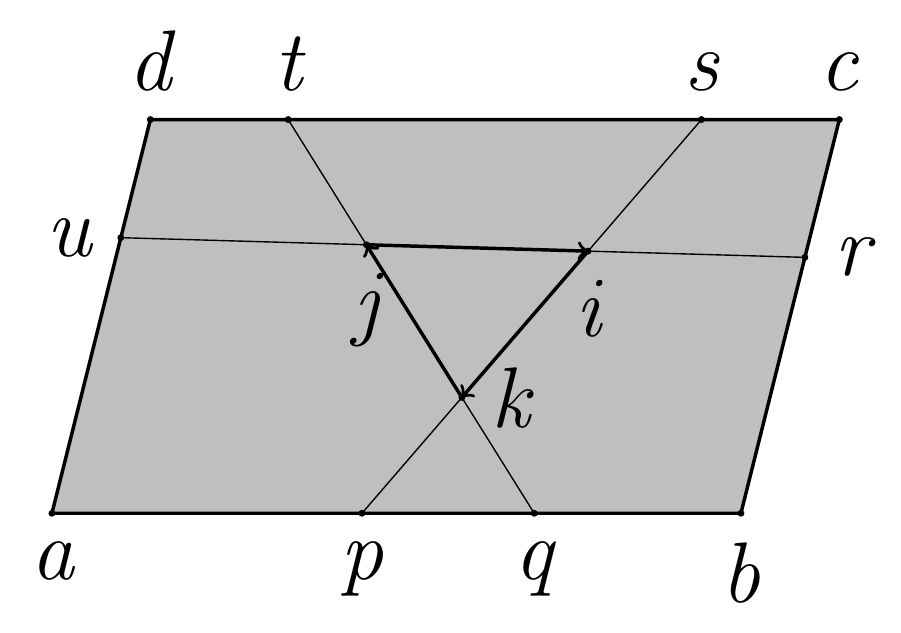}
\end{tabular}
\caption{Type~IV structure change.}\label{f-4}
\end{figure}

\subparagraph*{Structure changes} 
The subdivision of a contact facet by other contact facets is continuous in $t$,
except for four types of structure changes.  I.~Facets appear or disappear
at the bounds of their intervals.
II.~FF-edges appear or disappear when a facet vertex hits a facet or
when two facet edges intersect.  In Fig.~\ref{f-2}, the FF-edge $vw$ appears
when the facet vertex $a$ hits the facet $def$ or when the facet edges $ac$ and
$de$ intersect.  III.~FFF-vertices appear or disappear when a facet edge $e$
hits an FF-edge, and two EF-vertices swap position on $e$.  In Fig.~\ref{f-3},
the EF-vertices $j$ and $k$ swap on the facet edge $df$ when it hits the FF-edge
$gi$, and the FFF-vertex $x$ appears.  IV.~The FFF-vertices of four facets swap
position along their FF-edges when the facets intersect at a vertex.  In
Fig.~\ref{f-4}, three facets intersect $abcd$ in FF-edges $ps$, $qt$, and $ur$,
FFF-vertices $i$ and $j$ swap on $ur$, $j$ and $k$ swap on $qt$, and $i$ and $k$
swap on $ps$.

\subparagraph*{Structure change a-polys} Structure changes occur at zeros of
a-polys.  The Type~I facet interval a-polys are discussed above.
Fig.~\ref{f-types} describes the a-polys of the other types.

\begin{figure}[htbp]
\begin{tabular}{ll}
IIa & a) zero distance from a facet vertex to the plane of a facet.\\  
    & b) 3-contact of vertex, 1-contact of facet.\\ 
IIb & a) zero volume of the tetrahedron defined by two facet edges.\\  
   & b) 2-contact of each edge.\\
III & a) zero distance from an EF-vertex to the plane of a facet.\\ 
  & b) 2-contact of EF-vertex edge, 1-contact of EF-vertex facet, 1-contact of facet.\\
IV & a) zero distance between two FFF-vertices along an FF-edge.\\
  & b) 1-contact of each facet (four in all) incident on the FFF-vertices.
\end{tabular}
\caption{Structure change conditions (a) and a-polys (b).}\label{f-types}
\end{figure}

\subparagraph*{Sweep algorithm}
We construct the subdivision of a contact set by constructing the subdivision of
each contact facet at its initial $t$ value, sweeping along its $t$ interval,
computing the $t$ values where the subdivision undergoes structure changes, and
updating the structure.  The sweep state is the ordered list of EF-vertices
along each contact edge, the set of interior EF-vertices, the set of FF-edges,
and the ordered list of FFF-vertices along each FF-edge.  The sweep events are
the angles where 1) EF-vertices and 2) FF-edges appear or disappear (at a Type I
or II structure change) 3) an internal EF-vertex hits an FF-edge (III), 4) two
EF-vertices swap on a contact edge (III), or 5) two FFF vertices swap on an
FF-edge (IV).  FFF-vertices can appear and disappear at events 2, 3 or 4.
Events 1 and 2 are calculated before the sweep, event 3 is calculated at a 1 or
2 event, and events 4 and 5 are calculated when two vertices become adjacent on an
edge.  To calculate event 3, compute the zeros of its a-poly
(Fig.~\ref{f-types}III).  When sweeping through each zero, check if the FF-edge
and EF-vertex exist at that angle and evaluate predicates at that angle to
determine if the EF-vertex hits the line of the FF-edge between its current
endpoints, and if so, where in the list of FFF-vertices the new one should be
inserted.  Similarly for 4 and 5.

\subparagraph*{Handling identities} Each predicate has the same zero set as an a-poly.
For example, testing if a contact edge pierces a contact facet involves testing
an (endpoint) contact vertex vs. the contact facet.  The associated a-poly
contains the vertex 3-contact and the facet 1-contact.  Before evaluating a
predicate at an event angle, we check for identity.  An identity results from
evaluating the predicate at a parameter $t$ that is a zero of a $k\geq 1$
repeated irreducible factor of the a-poly of the predicate.  We replace the sign
of the predicate with the sign of its $k$th derivative, evaluated using
automatic differentiation.  The derivative gives the sign of the predicate value
immediately after the event, as required by the sweep algorithm.

\begin{figure}[htbp]
\begin{center}
\begin{tabular}{ccc}
\includegraphics[height=1.5in]{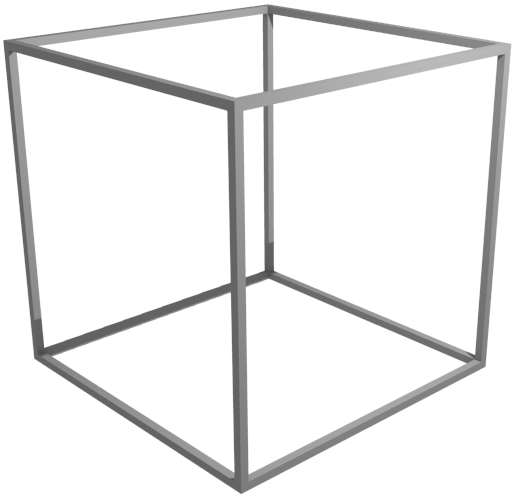} &
\includegraphics[height=1.5in]{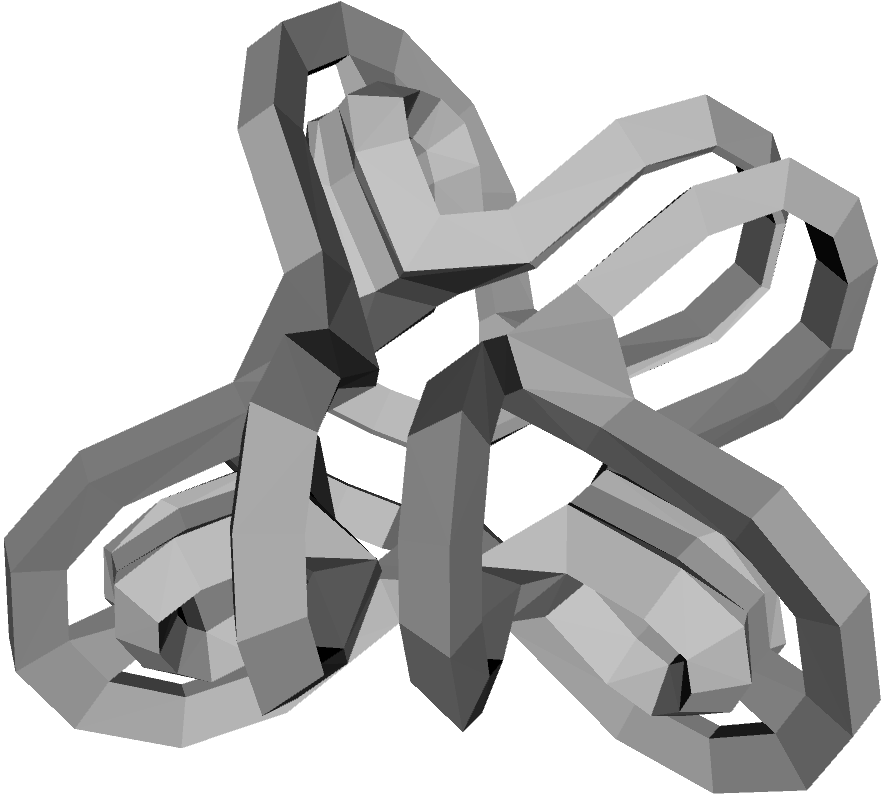} \\
frame: 40 vertices and 96 facets & knot: 480 vertices and 992 facets\\
\includegraphics[height=1.5in]{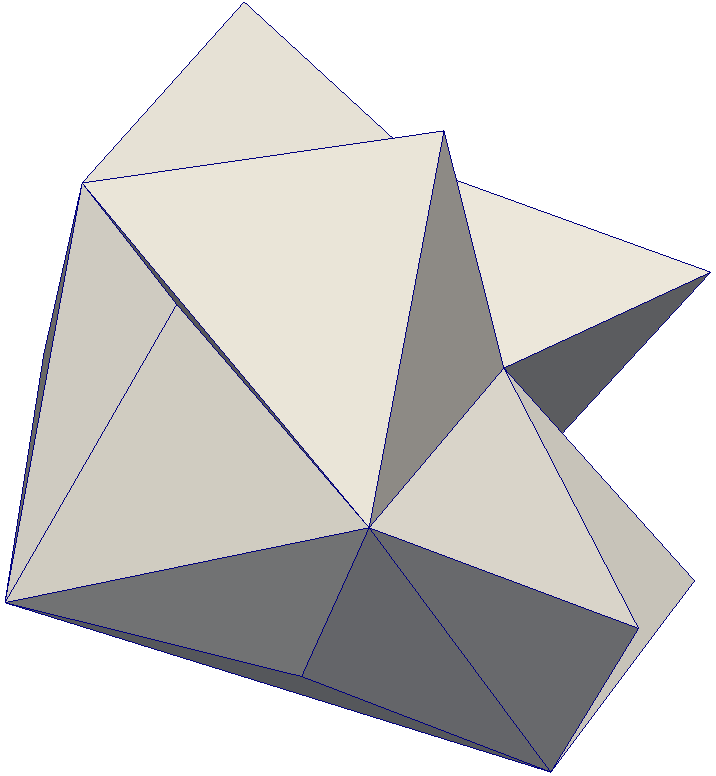} &
\includegraphics[height=1.5in]{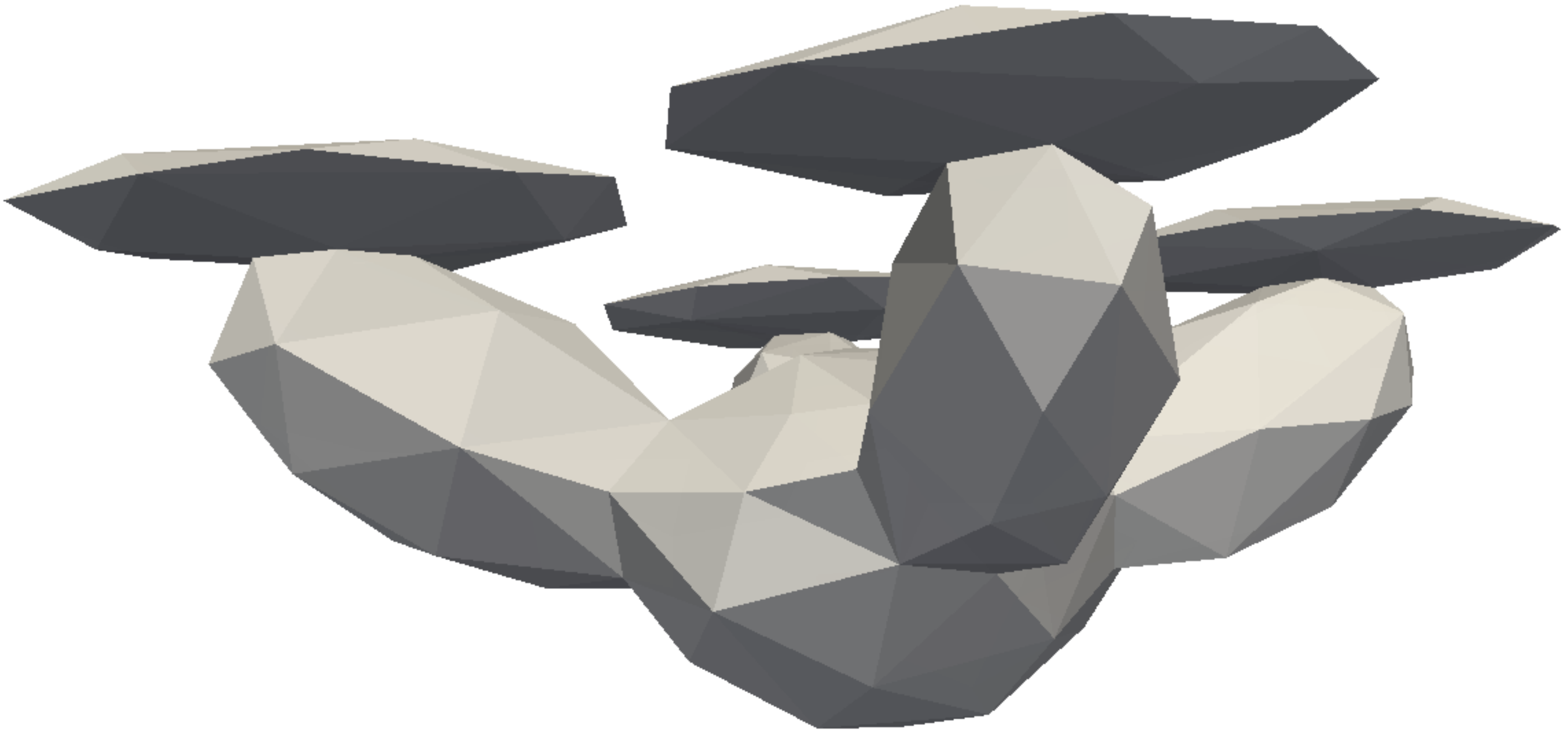} \\
simple: 18 vertices and 32 facets & drone: 189 vertices and 374 facets\\
\end{tabular}
\end{center}
\caption{Sweep Inputs.  Actual frame and knot are large enough for simple and
  drone to fly through.}
\label{f-inputs}
\end{figure}

\section{Results}\label{s-results}

We measure the running time of predicate evaluation using the factoring
algorithm (Sec.~\ref{s-factoring}) for identity detection and compare it to
using the greatest common divisor (GCD) for degeneracy detection.

We compute the GCD with Euclid's algorithm.  The main step is polynomial
division.  We compute the degree of a remainder by finding its nonzero
coefficient of highest degree.  If the sign of a coefficient is ambiguous in
double precision interval arithmetic, we evaluate modulo several 32-bit primes
and invoke the Chinese Remainder Theorem (CRT).  The a-polys have degree at most
9 in the input coordinates (Sec.~\ref{s-apoly}).  The leading coefficient is at
most degree~9 in the polynomial coefficients.  Hence, CRT requires
$\lceil{9\cdot 9 \cdot 53 / 32}\rceil=135$ modulo evaluations.  If they are all
zero, the coefficient is zero.  This analysis assumes that all inputs have the
same exponent field and can be replaced by their 53-bit integer mantissas.

In our first set of tests, we selected 100,000 representatives at random and
instantiated them on a pool of 12 $O$ vertices and 12 $R$ vertices with random
coordinates in $(-1,1)$.  We factored the univariates of these a-polys, isolated
the zeros of the factors, and stored them in a red-black tree.

Let $t_h$ be the $i$th largest real zero of $h$, a factor of a random a-poly
$g$.  When inserting $t_h$ into the tree, it must be compared to prior zeros,
such as the $j$th zero $t_e$ of $e$, a factor of $f$.  If $e=h$ and $i=j$, then
$t_h=t_e$.  To measure the running time of the identity detection algorithm
(Sec.~\ref{s-factoring}), we add an unnecessary test if $g(t_e)$ is an identity.
This ensures identity tests on random polynomials with both possible answers.
Adding these 9,660,759 identity tests (221,252 positive) increases the running
time from 6.6 seconds to 18.1 seconds, giving an average identity detection time
of 1.2 microseconds.

To test the GCD approach, we drop factoring and the equality test, and replace
each polynomial $f$ with its square-free form $f/{\mathrm{GCD}}(f,f')$.  If the
comparison of zeros $t_f$ of $f$ and $t_g$ of $g$ is ambiguous in double
precision, we run the following degeneracy test on $g(t_f)$.  Set $h\leftarrow
{\mathrm{GCD}}(f,g)$ and $e = f/h$.  If $e(t_f)$ is unambiguously nonzero,
$g(t_f)$ must be zero.  If $e(t_f)$ and $g(t_f)$ are both ambiguous, redo these
steps with more precision.  The additional time was 376 seconds for 81264
degeneracy tests, for an average time of 4627 microseconds.  To be sure that
$t_f=t_g$ and not some other zero of $g$, we must check that its comparison with
other zeros is unambiguous, so the true cost of the GCD method is even higher.

In the second set of tests, we ran a sweep algorithm (Sec.~\ref{s-sweep}) on the
polyhedra shown in Fig.~\ref{f-inputs}.  Table~\ref{t-sweep} shows the average
running times for sweeping a facet using factor-based identity detection and GCD
degeneracy detection.  We sweep either all the facets (tests~1-3) or a large
random sample over all angles (test~4).

\begin{table}[htbp]
\caption{Sweep algorithm: $f$ total number of contact
  facets, $t_{\mathrm{fac}}$ and $t_{\mathrm{GCD}}$ average running time in
  seconds for sweeping a facet with our identity detection and GCD-based
  degeneracy detection.}
  \label{t-sweep}
  \centering
  \begin{tabular}{lllrrrrr}
    & $O$ & $R$ & $f$ & $t_{\mathrm{fac}}$ & $t_{\mathrm{GCD}}$ & $t_{\mathrm{GCD}}/t_{\mathrm{fac}}$ \\
    1 & frame & simple & 2196 & 0.021 & 0.533 & 25.33 \\
    2 & frame & drone & 18812 & 0.148 & 2.093 & 14.13 \\
    3 & knot & simple & 23419 & 0.023 & 0.651 & 27.24 \\
    4 & knot & drone & 235789 & 0.037 & 0.857 & 23.26
  \end{tabular}
\end{table}

The first tests indicate that factor-based identity detection is
3500 times faster than GCD-based degeneracy detection.  The sweep tests
show that the effect of this improvement entails a factor of
14 speedup in sweep time.  Since identity detection is sped up by 3500,
factor-based identity detection uses less that 0.5\% of the
overall running time, versus 90\% for GCD-based degeneracy detection.

\section{Discussion}
\label{f-discuss}

We have shown that looking up the factors of an a-poly is much faster than
polynomial algebra for zero detection.  As an additional advantage,
factorization provides a unique representation of each algebraic number as the
$i$th zero of an irreducible polynomial.

In future work, we will extend the sweep algorithm to completely construct the
subdivision of a contact set.  We are missing the surfaces that bound the cells
and their nesting order.  We will construct a connected component of the free
space boundary by visiting the neighboring contact facets, computing their
subdivisions, and so on.  All
the predicates are angle polynomials evaluated at zeros of angle polynomials.

For efficient free space boundary construction, we must eliminate irrelevant
contact facets from consideration and must eliminate irrelevant sweep angles for
relevant contact facets.  One strategy is to construct a polyhedral inner and
outer approximation of $R$ as it sweeps through a small angle.  For that angle
range, the boundary of the rotational free space lies between the boundaries of
the translational free spaces of the approximations.  We can use our fast
polyhedral Minkowski sum software \cite{kyung-sacks-milenkovic15} to generate
approximations of the rotational free space boundary for a set of angle
intervals covering the unit circle.  We see no reason that sweeping a relevant
facet should have fewer identities than sweeping an irrelevant facet, and so
fast identity detection should provide the same speedup as observed in
Sec.~\ref{s-results}.

We conclude that our identity detection algorithm is useful for the
drone--warehouse problem.  But does the technique generalize to other domains?
We discuss three challenges.

Factor table construction (Sec.~\ref{s-offline}) depends on the property that
the set of polynomials is closed under factorization.  What if this is not true
for some alternate class of polynomials?  We would realize that something was
missing when factor table construction failed due to unmatched univariates.  We
would then analyze the failure to uncover the missing polynomials.  (This is how
we discovered the three-edge parallel-feature a-polys.)  The new polynomials
might also have had unmatched factors.  However, the process of adding missing
factor polynomials must converge because factoring reduces degree.

Representative generation (Sec.~\ref{s-factoring}) depends strongly on the
a-poly representation.  However, the approach should generalize.  A predicate
has symmetries on its inputs that leave it alone or flip its sign.  Two
invocations of a predicate function (with repeated inputs) are isomorphic if one
can get from one to the other by applying those symmetries and reindexing
inputs.  A representative is the isomorphism class member that is
lexicographically minimal.

Generalizing the matching of factors to prior classes for table generation would
greatly increase the running time because it uses enumeration of permutations.
In free space construction for $R$ with $d$ degrees of freedom, the
computational complexity is $(3d!)d^2$, which would balloon to 80,000 years for
$d=6$ (unconstrained rotation and translation).  We could use the ``Birthday
paradox'' to match two polynomials by generating the square root of the number
of permutations for each one and finding a collision, reducing the running time
by a factor of $\sqrt{3d!}$ to less than a day.  Moreover, these enumerations
can be tested in parallel.

Greater complexity might also increase the time required to construct a
representative (Sec.~\ref{s-factoring}) for identity detection, but this cost
can be reduced by using a subset of the symmetry group and increasing the size
of the lookup table.  For example, if we did not reorder the elements, the table
size would be 134,082, and if we did not reorder the left and right columns
either, it would be 972,806.  These tables are easy to generate from the minimal
table, and both are reasonable for current RAM sizes.

\bibliographystyle{plainurl}
\bibliography{bib,bib2}
\end{document}